\documentstyle[12pt,epsf]{article}
\begin{document}
\title{{\bf Featuring the  structure functions geometry}}
\author{
S. M. Troshin, N. E. Tyurin\\
\small \it
 Institute for High Energy Physics,\\
\small \it
 Protvino, 142284 Russia}
\normalsize
\date{}
\maketitle
\begin{abstract}
We consider geometrical properties of the polarized and unpolarized
structure functions and provide  definition for the
$b$--dependent structure functions. It is shown that unitarity does
not allow factorized form of the structure functions
over the $x$ and $b$ variables. We conclude that the spin of constituent
quark has a significant orbital angular momentum component.
\end{abstract}

\section*{Introduction}
The behavior and dependence of the  structure functions
on the Bjorken $x$ is among the most actively discussed
subjects in the unpolarized and polarized
deep-inelastic scattering. The particular role here
belongs to the small $x$ region where asymptotical properties
of the strong interactions can be studied. The characteristic
 point  of  low-$x$ region is an essential
nonperturbative nature of the underlying dynamics in the whole
region of $Q^2$\cite{bj,pasc}. Despite the results
of  perturbative QCD calculations  are in a
  good
agreement with the latest HERA data,
 the conceptual
 feasibility of the perturbative QCD methods
in this region has not been justified.

 Of course, the shortcomings of various model approaches
to the study of this nonperturbative region are also evident.
However, one can hope to gain from these models an information
which cannot  be obtained from the perturbative methods
(cf. \cite{bj}).
Among the possible extensions there could be  considerations
of the geometrical features of the structure functions, i.e.
dependence  of the structure functions on the transverse
coordinates or in  other words on impact parameter.
This dependence  would allow one to gain an
information on the spatial distribution of the partons inside
the parent hadron and the spin properties of the nonperturbative
intrinsic hadron structure. The geometrical properties of
structure functions should play an important role
 under analysis of
the lepton--nuclei deep--inelastic scattering and in the
hard production in the heavy--ion collisions.

\section{Definition and interpretation of  $b$--dependent
structure functions}

In this note we study the $b$--dependence of the structure functions
along the line used in \cite{usdif}, i.e. we suppose that the
deep--inelastic scattering is determined by the aligned-jet
mechanism \cite{bj}. There are serious arguments in favor of
its leading role and dominance over the other mechanism known
as a color-transparency. The aligned-jet mechanism
  is  essentially nonperturbative
one and allows to relate structure functions with the
discontinuities of the amplitudes
 of quark--hadron elastic scattering.
These relations have the following form \cite{jj,sf}
 \begin{eqnarray}
q(x) & = & \frac{1}{2}\mbox{Im}[F_1(s,t)+F_3(s,t)]|_{t=0},\nonumber \\
\Delta q(x) & = & \frac{1}{2}\mbox{Im}[F_3(s,t)-F_1(s,t)]|_{t=0},\nonumber \\
 \label{def1}
\delta q(x) & = & \frac{1}{2}\mbox{Im} F_2(s,t)|_{t=0}.
\end{eqnarray}
 The functions $F_i$ are  helicity amplitudes
for the elastic quark-hadron scattering in the standard
notations for the nucleon--nucleon scattering.
We consider high energy limit or the region of small
$x$.

The structure functions obtained according to the above
formulas should be multiplied by the factor $\sim 1/Q^2$
-- probability that such aligned--jet configuration
 occurs \cite{bj}.

The amplitudes $F_i(s,t)$ are the corresponding Fourier-Bessel
transforms of the functions $F_i(s,b)$.

The relations Eqs. (\ref{def1}) will be used as a starting
point under definition of the structure functions which
depend on impact parameter. According to these relations it
is natural to give the following operational  definition:
\begin{eqnarray}
q(x,b) & \equiv & \frac{1}{2}\mbox{Im}[F_1(x,b)+F_3(x,b)],\nonumber \\
\Delta q(x,b) & \equiv & \frac{1}{2}\mbox{Im}[F_3(x,b)-F_1(x,b)],\nonumber \\
\delta q(x,b) & \equiv & \frac{1}{2}\mbox{Im}F_2(x,b),\label{def2}
\end{eqnarray}
and $q(x)$, $\Delta q(x)$ and $\delta q(x)$ are the integrals
over $b$ of the corresponding $b$-dependent distributions, i.e.
\begin{equation}\label{int1}
q(x)=\frac{Q^2}{\pi^2 x}\int_0^\infty bdb q(x,b),\quad
\Delta q(x)=\frac{Q^2}{\pi^2 x}\int_0^\infty bdb \Delta q(x,b)
\end{equation}
and
\begin{equation}\label{int2}
\delta q(x)=\frac{Q^2}{\pi^2 x}\int_0^\infty bdb \delta q(x,b).
\end{equation}

The functions $q(x,b)$, $\Delta q(x,b)$ and $\delta q(x,b)$ depend
also on the variable $Q^2$  and have  simple  interpretations, e.g.
the function $q(x,b,Q^2)$ represent probability to find  quark $q$ in the
hadron with a fraction of its longitudinal momenta $x$ at the transverse
distance
\[
b\pm \Delta b,\quad
 \Delta b\sim 1/Q
\]
 from the hadron geometrical center.
 Interpretation of the spin distributions directly follows
from their definitions:  they are the differences of the probabilities to find
quarks in the two spin states with longitudinal or transverse
directions of the quark and hadron spins.

It should be noted that the unitarity plays crucial role
in the direct probabilistic
interpretation of the function $q(x,b)$. Indeed due to unitarity
\begin{equation}\label{un1}
0\leq q(x,b)\leq 1.
\end{equation}
The integral $q(x)$ is a quark number density which
is not limited by unity and can have arbitrary nonnegative
value.
Thus, the given definition of the $b$--dependent structure
functions is self-consistent.  Of course,
spin distributions $\Delta q(x,b)$ and $\delta q(x,b)$ are not positively
defined.

\section{Unitarity and structure function geometrical profiles}

The unitarity can be fulfilled through the $U$--matrix representation
for the helicity amplitudes of elastic quark--hadron scattering.
In the impact parameter representation the expressions for the
helicity amplitudes are the following \cite{usdif}
\begin{eqnarray}
& & F_{1,3}(x,b)  =  U_{1,3}(x,b)/[1-iU_{1,3}(x,b)],\nonumber \\
& & F_2(x,b)  =  U_2(x,b)/[1-iU_1(x,b)]^2\label{fu}
\end{eqnarray}
Unitarity requires Im$U_{1,3}(x,b)\geq 0$. The $U$--matrix
 form of unitary
representation contrary to the eikonal one does not generate
itself essential singularity in the complex $x$ plane
 at $x\to 0$ and   implementation of unitarity can be performed
easily. Therefore we use this representation and
  not the method of the
eikonalization.
The model which provides explicit form of helicity functions
$U_i(x,b)$ has been described elsewhere \cite{usdif}.
A hadron consists of the constituent quarks aligned in the
longitudinal direction and embedded into the nonperturbative
vacuum (condensate). The constituent quark appears
as a quasiparticle, i.e. as current valence quark surrounded by
the cloud of quark-antiquark pairs of different flavors. The strong
interaction radius of the constituent quark $Q$ is determined
by its Compton wavelength.

Spin of constituent quark, e.g. $U$-quark in this approach is given
by the  sum:
\begin{equation}\label{spb}
J_U=1/2=S_{u_v}+S_{\{\bar q q\}}+L_{\{\bar q q\}}=
1/2+S_{\{\bar q q\}}+L_{\{\bar q q\}}.
\end{equation}
In the model an exact compensation between the total spin
of the quark-antiquark cloud and its
 angular orbital momenta occurs, i.e.
\begin{equation}\label{comp}
L_{\{\bar q q\}}=-S_{\{\bar q q\}}.
\end{equation}
In this approach based on effective Lagrangian the gluon
degrees of freedom are overintegrated.

On the grounds of the experimental data for polarized DIS
we arrive to conclusion that the significant part of the spin
of constituent quark is due to the orbital angular momentum
of the current quarks inside the constituent one \cite{spcon}.

The explicit expressions for the helicity functions $U_i(x,b)$
at small $x$ can be obtained from the corresponding functions
$U_i(s,b)$ given in \cite{usdif} by the substitute $s\simeq Q^2/x$
and at small values of $x$ they get the form:
\begin{eqnarray}
& &U_{1,3}(x,b)=U_0(x,b)[1+\beta _{1,3}(Q^2)m_Q \sqrt{x}/Q],\nonumber\\
& &U_2(x,b)=g_f^2(Q^2)\frac{m_Q^2x}{Q^2}\exp[-2(\alpha -1)m_Qb/\xi]U_0(x,b),
\label{uis}
\end{eqnarray}
where
\begin{equation}\label{u0}
U_0(x,b)=i\tilde U_0(x,b)
=i\left[\frac{a(Q^2)Q}{m_Q\sqrt{x}}\right]^{n+1}\exp[-Mb/\xi].
\end{equation}
  $a$, $\alpha$, $\beta$, $g_f$ and $\xi$ are the model
parameters, some of them in this particular case of quark-hadron
scattering depend on the virtuality $Q^2$.The
 meaning of these parameters (cf. \cite{usdif}) is not
crucial here; note only that $m_Q$ is
the average mass of constituent quarks in the quark-hadron system
of $n+1$ quarks and $M$ is their total mass, i.e.
$M=\sum_{i=1}^{n+1}m_i$. We  consider here for simplicity
a pure imaginary amplitude.  Using definition of the $b$--dependent structure
functions given in  Sec. 1 and Eqs. (\ref{fu}) we obtain at small
$x$:
\begin{equation}\label{qxb}
q(x,b)=\frac{\tilde U_0(x,b)}{1+\tilde U_0(x,b)},
\end{equation}
\begin{equation}\label{dqxb}
\Delta q(x,b)=\frac{\beta_-(Q^2) m_Q\sqrt{x}}{Q}\frac{\tilde U_0(x,b)}
{[1+\tilde U_0(x,b)]^2},
\end{equation}
\begin{equation}\label{dsqxb}
\delta q(x,b)=\frac{g_f^2(Q^2)m_Q^2x}{Q^2}
\exp[-\frac{ 2(\alpha -1)m_Qb}{\xi}]
\frac{\tilde U_0(x,b)}{[1+\tilde U_0(x,b)]^2},
\end{equation}
where $\beta_-(Q^2)=\beta_3(Q^2)-\beta_1(Q^2)$.
From the above expressions it follows that $q(x,b)$ has a central
$b$--dependence, while $\Delta q(x,b)$ and $\delta q(x,b)$ have
 peripheral profiles. Their qualitative dependence on the impact parameter
$b$ is depicted in Fig. 1. The peripheral dependence on impact
parameter according to the relation Eq. (\ref{comp}) is the
manifestation of a significant presence of the angular orbital
momenta in the spin balance of a nucleon.
\begin{figure}[hbt]
 \hspace*{3cm}
 \epsfxsize=80  mm  \epsfbox{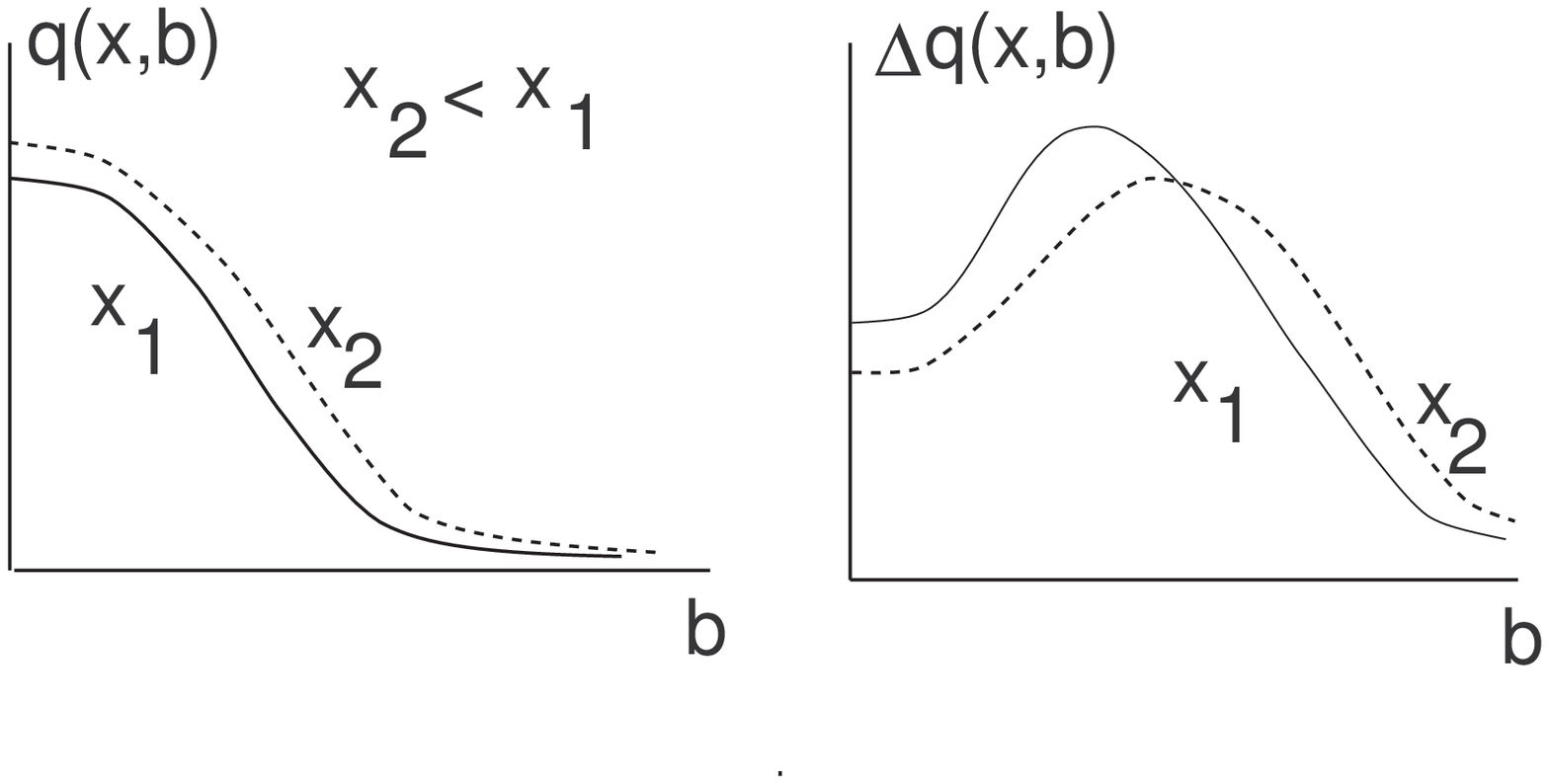}
\caption[junk]{{\it
Low-$x$ dependence on $b$ of the  structure functions
 $q(x,b)$ and $\Delta q(x,b)$}}

 \end{figure}

From Eqs.(\ref{qxb}--\ref{dsqxb}) it follows that factorization
of $x$ and $b$ dependencies is not allowed by unitarity.
However, this result is valid for the small $x$ region only and
approximate factorization is possible in the region of  not
too small $x$ where account of unitarity  reduces to the
factorization breaking corrections.

The following relation between the structure functions $\Delta q(x,b)$
and $\delta q(x,b)$ can also be inferred from the above formulas
\begin{equation}\label{rel}
\delta q(x,b)=c(Q^2)\frac{\sqrt{x}}{Q}\exp(-\gamma b)\Delta q(x,b).
\end{equation}
Thus, the function $\delta q(x,b)$ which describes  transverse
spin distribution is suppressed by the factors
$\sqrt{x}$ and $\exp(-\gamma b)$, i.e. it has a more central
profile. This suppression also reduces double-spin transverse
asymmetries in the central region in the Drell-Yan production
compared to the corresponding longitudinal asymmetries.

The strange quark structure functions have also a more central
$b$--dependence than in the case of $u$ and $d$ quarks.
 The radius of the
corresponding quark matter distribution is
\begin{equation}\label{rad}
R_q(x)\simeq \frac{1}{M}\ln{Q^2}/{x}
\end{equation}
 and the ratio of the strange quark distributions to the
light quark distributions radii is given by the corresponding
constituent quark masses,
i.e. for the nucleon this ratio would be
\begin{equation}\label{ratr}
R_s(x)/R_q(x)\simeq (1+\frac{\Delta m}{4m_Q})^{-1},
\end{equation}
where $\Delta m = m_S-m_Q$.

Time reversal invariance of strong interactions allows
one to
write down relations similar to Eqs.(\ref{def1}) for the
fragmentation functions also and obtain expressions
for the fragmentation functions $D_q^h(z,b)$, $\Delta D_q^h(z,b)$,
$\delta D_q^h(z,b)$ which have just the same dependence
on the impact parameter $b$
as the corresponding structure functions. The fragmentation
function $D_q^h(z,b,Q^2)$ is the probability for fragmentation
of  quark $q$ at  transverse distance
 $b ~ \pm ~ \Delta~ b$ ($\Delta b\sim 1/Q$)
into a hadron $h$ which carry the fraction
$z$ of the quark momentum. In this case $b$ is a transverse
distance between quark $q$ and the center of the
hadron $h$. It is positively
defined and due to unitarity obey to the inequality
\begin{equation}
0\leq D_q^h(z,b)\leq 1
\end{equation}
The physical interpretations of  spin--dependent
 fragmentation functions $\Delta D_q^h(x,b)$ and
$\delta D_q^h(x,b)$ is similar to that of  corresponding
spin structure function. Peripherality of the spin fragmentation
functions can also be considered as a manifestation
of the important role of angular orbital momenta.
\section*{Conclusion}
 It is interesting to note that the spin structure functions
have a peripheral dependence on the impact parameter contrary
to  central profile of the unpolarized structure function.
In the considered model where the hadron has aligned structure
the peripherality of the spin structure functions implies that
the main contribution to the spin of constituent quark is due to
the orbital angular momentum.
This orbital angular momentum has a nonperturbative origin
and does not result from the perturbative QCD evolution.
This conclusion  provides clue for the possible solution of the
problem of the nucleon spin structure. It is interesting
to find out possible experimental signatures of the peripheral
geometrical profiles of the spin structure functions and
the significant role of the orbital angular  momentum.
One of such indications could be an observation of the
different spatial distributions of charge and magnetization
at Jefferson Lab \cite{bur}. It would also be important to
have a precise data for the strange formfactor.
It could also be done analyzing both the spin structure and
 angular distribution in exclusive electroproduction and
it  worth considering in a separate study.
\section*{Acknowledgements}
This work was supported in part by the Russian Foundation for
Basic Research under Grant No. 99-02-17995. One of the authors
(S.T.) is grateful to Alan Krisch for the warm hospitality at
the Spin Physics Center of the University of Michigan where
this work was finished.

\end{document}